%% file: main.tex
\documentclass[lettersize,journal]{IEEEtran}
\usepackage{template}
\usetikzlibrary{arrows,automata}
\begin{document}
\pdfimageresolution=300
\title{Coded Estimation: Design of  Backscatter Array Codes for 3D Orientation Estimation}
\author{%
  \IEEEauthorblockN{M. Rida Rammal$^\star$, Suhas N. Diggavi$^\star$, and Ashutosh Sabharwal$^\dagger$}\\
  \IEEEauthorblockA{$^\star$Department of Electrical and Computer Engineering, UCLA. \\
                    $^\dagger$Department of Electrical and Computer Engineering, Rice University.}
                    }

\maketitle
\begin{abstract}
We consider the problem of estimating the orientation of a 3D object with the assistance of configurable backscatter tags. We explore the idea of designing tag response codes to improve the accuracy of orientation estimation. To minimize the difference between the true and estimated orientation, we propose two code design criteria. We also derive a lower bound on the worst-case error using Le Cam's method and provide simulation results for multiple scenarios including line-of-sight only and multipath, comparing the theoretical bounds to those achieved by the designs.
\end{abstract}

\section{Introduction}
3D orientation tracking is an important function in many domains, e.g., in robotics, aerospace, and medicine. Orientation can be estimated using many methods, e.g., inertial sensors can be mounted on the object whose orientation has to be measured.  Orientation can also be measured using a computer vision based methods\cite{ 5152855, 5152739, inbook, article}. Each of the existing methods have their own pros and cons. For example, inertial sensors may not be suitable for many Internet-of-Things applications, e.g., tracking packages where the solutions have to be very  cost-effective. Additionally, the performance of computer vision methods  methods depends on light conditions, and can be confused for objects that exhibit symmetries. As an extreme example of that
last problem, one can rotate a cube in many ways such that it would look precisely the same as it did prior to rotation.


Wireless sensing has emerged as an interesting alternative sensing modality. Radio-frequency based methods can be useful when visible light wavelengths are not effective, e.g., cases with poor visibility or non-line-of-sight scenarios. In particular, backscatter arrays have been recently used for geo-location and 3D orientation estimation\cite{10.1145/3447993.3448627, 10.1145/3448082, wei2016gyro, shirehjini2012rfid}. Using backscatter arrays is philosophically akin to ``painting the faces" of an object, making it a promising option for the orientation detection of symmetric objects such as the solid cube in the last example. In this work, we study 3D orientation estimation with the help of configurable backscatter arrays.

In order to aid with the estimation task, one can design the backscatter response to received signals. Specifically, we design the backscatter responses by changing their reflectivities. The design can be captured as a binary code specifying what or when the backscatter tags reflect. The problem of finding the best code for the estimation task was first formulated and explored in \cite{CRSD-ISIT20}, where we proposed a heuristic code design criterion and had a preliminary exploration of the systems performance. 

\noindent \textbf{Contributions:} In this paper, we revisit the problem and expand our understanding in the following ways. First, we propose two analytic code design criteria that depend on channel knowledge, and investigate their performance with respect to a baseline orthogonal code. We also develop a lower bound on the worst-case error, which quantifies the systems performance with respect to channel parameters such as the number of antennas and the number of tags used. Finally, we provide a comprehensive numerical exploration of the systems performance, including the impact of multipath, and the robustness of the design criteria against imperfect channel knowledge.

By numerically evaluating the performance, we observe the following. Our design criteria yield codes that offer significant advantages over the channel-oblivious orthogonal code for both the average and worst-case error performance (see Section \ref{sec:perf_criteria} for precise definitions). In some cases, our designs provide two orders of magnitude improvements in error performance. We also found that \emph{precise} channel knowledge at the transmitter is not critical; Our design criteria are robust against channel estimation errors, retaining almost all improvements when computed with SNR values greater than or equal to $10\mathrm{dB}$.  Furthermore, our numerical results suggest that multipath offers SNR gain but not multipath-diversity gain; this means that when one fixes the total energy received across paths, the performance does not improve when multipath is present.

\emph{Related work:} In \cite{CRSD-ISIT20}, we suggested a heuristic design criteria aimed at maximizing the average-case performance, but did not make attempts to assess its performance. Furthermore, we only considered the average-case error of the system, and did not study the worst-case error, explored in this paper. In ~\cite{wei2016gyro,shirehjini2012rfid,jiang2018orientation}, the authors present various methods for orientation estimation (2D or 3D) using backscatter tags, but do not explore the design of backscatter responses and its effect on the estimation problem. In \cite{fu2017simultaneous}, the authors use orthogonal codes (see Section \ref{sec:numerics} for a definition) to distinguish the tags from each other and the environment. However, none of these past works has considered optimal code design for 3D orientation sensing. 

\emph{Organization:} Section~\ref{sec:system} describes the channel model and formulates the estimation problem.  Section~\ref{sec:results} describes the main results, including the two code design criteria. Section~\ref{sec:numerics} provides the various numerical results. Finally, Section~\ref{sec:proofs} provides proofs for the theorems and lemmas shown in Section~\ref{sec:results}.
\section{Problem Formulation}
\label{sec:system}
\begin{figure}[t!]
\centering
\begin{circuitikz}[scale=1.25]
\node [txantenna, scale =0.25] at (-3,0.25) {};
\node[text width=0cm] at (-2.7,0) {\vdots};
\node [rxantenna, scale=0.25] at (-3,-1) {};
\draw (-1,0) arc (180:360:1cm and 0.5cm);
\draw[dashed] (-1,0) arc (180:0:1cm and 0.5cm);
\draw (0,1) arc (90:270:0.5cm and 1cm);
\draw[dashed] (0,1) arc (90:-90:0.5cm and 1cm);
\draw (0,0) circle (1cm);
\shade[ball color=blue!10!white,opacity=0.20] (0,0) circle (1cm);
\node[draw, fill=red, star,star points=7,star point ratio=0.8, inner sep=1.5] at (-0.44,-0.44) {};
\node[draw, fill=red, star,star points=7,star point ratio=0.8, inner sep=1.5] at (0.44,0.44) {};
\node[draw, fill=red, star,star points=7,star point ratio=0.8, inner sep=1.5] at (-1,0) {};
\node[draw, fill=red, star,star points=7,star point ratio=0.8, inner sep=1.5] at (0,-1) {};
\draw (2 - \XA,\XB) arc (180:360:0.125 cm and 0.0625 cm);
\draw[densely dotted] (2 - \XA,\XB) arc (180:0:0.125 cm and 0.0625 cm);
\draw (2,0.125+\XB) arc (90:270:0.0625 cm and 0.125 cm);
\draw[densely dotted] (2,0.125+\XB) arc (90:-90:0.0625 cm and 0.125 cm);
\draw (2,\XB) circle (\XA cm);
\shade[ball color=blue!10!white,opacity=0.20] (2,\XB) circle (\XA cm);
\node[draw, fill=red, star,star points=7,star point ratio=0.8, inner sep=1.5] at (2, - 0.375+\XB) {};
\node[antenna, scale=0.125] at (2, -0.925+\XB){};
\node[rectangle,
draw = black,
minimum width = 1.75cm, 
minimum height = 1.5cm] (r) at (2.5,0.37) {};
\node[] at (2.65, 0.75) {\scriptsize Object};
\node[] at (2.55, 0.35) {\scriptsize Tag};
\node[] at (2.75, -0.025) {\scriptsize Antenna};
\end{circuitikz}
\caption{The components of the considered system. The object along with the tags, rotate around a specific point (in this case, the center of the sphere). The full-duplex antennas interrogate the configurable tags and receive back the reflected signal. We then use the received signal to infer the orientation.}
\label{fig:sys_comp}
\end{figure}
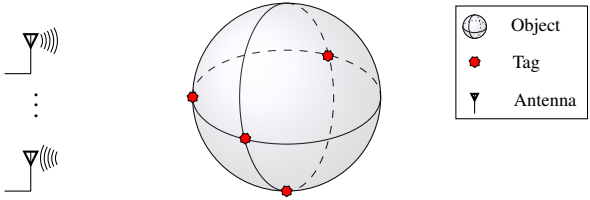
\subsection{Notation}
$\mathbb{R}^k$ and $\mathbb{C}^k$ are the sets of real-valued and complex-valued $k-$dimensional vectors, respectively. $A^T$ and $A^H$ denote the transpose and conjugate transpose of the matrix $A$. If $\mathbf{x}$ is a vector with $n$ elements, then $\textrm{diag}(\mathbf{x})$ denotes the $n\times n$ square matrix with the elements of $\mathbf{x}$ on its main diagonal;   $\left\lVert \cdot \right\rVert$ is the standard Euclidean norm on vectors, and $\left\lVert \cdot \right\rVert_F$ is the Frobenius norm on matrices.

\subsection{System Model}
We assume a system having $K$ full-duplex antennas with position vectors 
$\mathbf{x}_1^{\textrm{ant}}, \ldots,\mathbf{x}_K^{\textrm{ant}}\in\mathbb{R}^3$, and $N$ backscatter tags with position vectors $\mathbf{x}_1^{\textrm{tag}},\ldots, \mathbf{x}_N^{\textrm{tag}}\in\mathbb{R}^{3}$. We consider an object on which we place the $N$ configurable backscatter tags. The object can freely rotate around a specific point in space. For ease of computation, we consider the coordinate system in which the point that the object rotates around is the center $\textbf{0} = [0, 0, 0]^T$ (see Figure \ref{fig:sys_comp}).

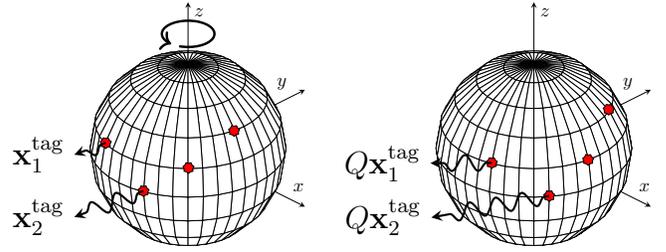
\begin{figure}[t!]
\centering
\tdplotsetmaincoords{60}{110}
\vspace{-0.65cm}
\begin{tikzpicture}[tdplot_main_coords, scale=0.7]
\begin{groupplot}[%
    group style={group size= 2 by 1, vertical sep=0cm, horizontal sep = -1 cm },
    height=5cm,width=6.4cm,
    ]

\nextgroupplot[axis equal,
        width=10cm,
        height=10cm,
        axis lines = center,
        xlabel = {$x$},
        ylabel = {$y$},
        zlabel = {$z$},
        xmin = -0.65, 
        xmax = 1.15,
        ymin = -0.65, 
        ymax = 1.15,
        zmin = -0.55,
        zmax = 1.2,
        ticks=none,
        enlargelimits=0.3,
        view/h=45,
        scale uniformly strategy=units only,]
\addplot3[%
        opacity = 1,
        surf,
        z buffer = sort,
        samples = 21,
        variable = \u,
        variable y = \v,
        domain = 0:180,
        y domain = 0:360,
        color=white,
        faceted color=black,
    ]
    ({cos(u)*sin(v)}, {sin(u)*sin(v)}, {cos(v)});
\nextgroupplot[axis equal,
        width=10cm,
        height=10cm,
        axis lines = center,
        xlabel = {$x$},
        ylabel = {$y$},
        zlabel = {$z$},
        xmin = -0.65, 
        xmax = 1.15,
        ymin = -0.65, 
        ymax = 1.15,
        zmin = -0.55,
        zmax = 1.2,
        ticks=none,
        enlargelimits=0.3,
        view/h=45,
        scale uniformly strategy=units only,]
 \addplot3[%
        opacity = 1,
        surf,
        z buffer = sort,
        samples = 21,
        variable = \u,
        variable y = \v,
        domain = 0:180,
        y domain = 0:360,
        color=white,
        faceted color=black,
    ]
    ({cos(u)*sin(v)}, {sin(u)*sin(v)}, {cos(v)});
\end{groupplot}
\node[draw, fill=red, star,star points=7,star point ratio=0.8, inner sep=1.5] at (3.78,3.04, 6.95) {};
    \node[draw, fill=red, star,star points=7,star point ratio=0.8, inner sep=1.5] at (2.81,3.45, 5.45) {};
    \node[draw, fill=red, star,star points=7,star point ratio=0.8, inner sep=1.5] at (4.57,5, 7.22) {};
    \node[draw, fill=red, star,star points=7,star point ratio=0.8, inner sep=1.5] at (3,5.35, 7.25){};
    \tdplotdrawarc[->,color=black, line width=0.32mm]{(0,3.33,7.35)}{0.5}{0}{340}{anchor=south west,color=black}{};
      \draw[thick, -stealth,decorate,decoration={snake,amplitude=3pt,pre length=2pt,post length=3pt}] (3.78,3.04, 6.95) -- ++(-1,-1,-1) node [font=\fontsize{3pt}{3pt}\selectfont,left]{\large $\mathbf{x}_1^{\mathrm{tag}}$};
    \draw[thick, -stealth,decorate,decoration={snake,amplitude=3pt,pre length=2pt,post length=3pt}] (2.81,3.45, 5.45) -- ++(-2.2,-2.2,-2.2) node [font=\fontsize{3pt}{3pt}\selectfont, left]{\large $\mathbf{x}_2^{\mathrm{tag}}$};

    \node[draw, fill=red, star,star points=7,star point ratio=0.8, inner sep=1.5] at (3.78 +\YA,3.04 +\YB, 6.95 + \YC) {};
     \node[draw, fill=red, star,star points=7,star point ratio=0.8, inner sep=1.5] at (-1,11.05, 5.57) {};
    \node[draw, fill=red, star,star points=7,star point ratio=0.8, inner sep=1.5] at (1.57,11.2, 6.2) {};
    \node[draw, fill=red, star,star points=7,star point ratio=0.8, inner sep=1.5] at (0,11.83, 7.37){};
    
    \draw[-stealth,decorate,decoration={snake,amplitude=3pt,pre length=2pt,post length=3pt}, thick]  (3.78 +\YA,3.04 +\YB, 6.95 + \YC) -- ++(-1.95,-1.95,-1.45) node [ left]{\large $Q\mathbf{x}_1^{\mathrm{tag}}$};
    \draw[-stealth,decorate,decoration={snake,amplitude=3pt,pre length=2pt,post length=3pt}, thick] (1.57,11.2, 6.2) -- ++(-3.75,-3.75,-3.25) node  [left]{\large $Q\mathbf{x}_2^{\mathrm{tag}}$};
\end{tikzpicture}
\caption{Rotation of the object about the $z$-axis. When the object rotates, the tags rotate along with it. The new positions of the tags produce a different received signal. While the tags are equidistant from the point of rotation in the figure, this does not have to be the case.}
\label{fig:sphere_rot}
\end{figure}

We specify the orientation of the object using a rotation matrix $Q\in SO(3)\subseteq \mathbb{R}^{3\times 3}$, where $SO(3)$ is the 3D rotation group. If the object has tags with positions $\mathbf{x}_1^{\textrm{tag}},\ldots, \mathbf{x}_N^{\textrm{tag}}$ on it, and a rotation $Q$ is applied to the object, the tags would have the new positions $Q\mathbf{x}_1^{\textrm{tag}},\ldots, Q\mathbf{x}_N^{\textrm{tag}}$ (see Figure \ref{fig:sphere_rot}). In this framework, we specify the original orientation of the object using the $3\times 3$ identity matrix $I_3$.

Each backscatter tag can be configured by setting its state to a value $i\in \{0,1\}$. The state, in turn, determines the reflectivity of the tag. For instance, we may choose to correspond a state of $0$ to a reflectivity of $+0.5$ and a state of $1$ to a reflectivity of $-0.5$. In this scenario, tags reflect regardless of the their state. On the other hand, we may choose to correspond a state of $0$ to a reflectivity of $0$ (does not reflect), and a state of $1$ to a reflectivity of $1$ (reflects) i.e. the state determines the on-off condition of the tag. In addition, the state of each tag can either remain constant (passive) or change over time (active). We study the performance of both options.

The orientation of the object determines the position of the tags, and the states of the tags set their reflectivities. Hence, the orientation and tag states determines the channel model, and in turn the received signal. We use the received signal to estimate the orientation of the object. Our goal is to analyze the performance of this system, and find the best set of tag states for the estimation task. We describe our channel model next. 
We use the following free-space path loss formula for line-of-sight propagation\cite{10.5555/1111206} between points $\mathbf{x}_1,\mathbf{x}_2\in\mathbb{R}^3$:
\begin{equation}
\label{eq:free_loss}
\eta(\mathbf{x}_1,\mathbf{x}_2) =  \frac{1}{4\pi\left\lVert \mathbf{x}_1-\mathbf{x}_2\right\rVert}\textrm{exp}\left(\frac{-2\pi j \left\lVert \mathbf{x}_1-\mathbf{x}_2\right\rVert}{\lambda}\right),
\end{equation}
where $\lambda$ is the wavelength. In other words, suppose $\mathbf{x}_{\mathrm{T_x}}$ and $\mathbf{x}_{\mathrm{R_x}}$ are the position vectors of an isolated transmitter and receiver, and $s_{\mathrm{T_x}}$ and $s_{\mathrm{R_x}}$ the transmitted and received signals, respectively, then 
\begin{equation}
\label{eq:free_loss_example}
s_{\mathrm{R_x}} = s_{\mathrm{T_x}}\eta(\mathbf{x}_{\mathrm{T_x}}, \mathbf{x}_{\mathrm{R_x}}).   
\end{equation}

We now specify the matrices involved in our channel model. Let $H_Q$ be the $K\times N$ matrix of the line-of-sight responses between the tags and antennas when the object is in orientation $Q$, i.e. $(H_Q)_{k,n} = \eta(\mathbf{x}_k^{\textrm{ant}}, Q\mathbf{x}_n^{\textrm{tag}})$. Let $A$ be the symmetric $K\times K$ matrix of inter-antenna line-of-sight responses with $A_{k,k'} = \eta(\mathbf{x}_k^{\textrm{ant}},\mathbf{x}_{k'}^{\textrm{ant}})$. Let $B$ be the symmetric $N\times N$ matrix of the inter-tag channel responses when the object is in orientation $Q$. In other words,
\begin{equation}
(B_Q)_{n,n'} = \eta(Q\mathbf{x}_n^{\textrm{tag}},Q\mathbf{x}_{n'}^{\textrm{tag}}) = \eta(\mathbf{x}_n^{\textrm{tag}},\mathbf{x}_{n'}^{\textrm{tag}}).
\end{equation}

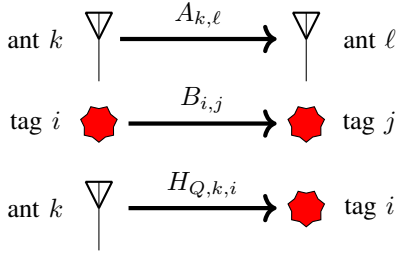
\begin{figure}[t!]
\centering
\begin{tikzpicture}[scale=1.125]
\node[bareantenna, scale=0.5] at (0,-1) {};
\draw[] (0,-1)--(0,-1.5);
\draw[->, ultra thick] (0.25,-1)  -- (2.1,-1) node[midway, above] {$A_{k,\ell}$};
\node[bareantenna, scale=0.5] at (2.45,-1) {};
\draw[] (2.45,-1)--(2.45,-1.5);
\node[] at (-0.75,-1) {ant $k$};
\node[] at (3.2,-1) {ant $\ell$};
\node[draw, fill=red, star,star points=7,star point ratio=0.8, inner sep=5] at (0,0+\ZA) {};
\node[] at (-0.75,0+\ZA) {tag $i$};
\draw[->, ultra thick] (0.35,0+\ZA)  -- (2.1,0+\ZA) node[midway, above] {$B_{i,j}$};
\node[draw, fill=red, star,star points=7,star point ratio=0.8, inner sep=5] at (2.45,0+\ZA) {};
\node[] at (3.2,0+\ZA) {tag $j$};
\node[bareantenna, scale=0.5] at (0,0+\ZB) {};
\draw[] (0,\ZB)--(0,-0.5+\ZB);
\node[] at (-0.75,0+\ZB) {ant $k$};
\draw[->, ultra thick] (0.35,0+\ZB)  -- (2.1,0+\ZB) node[midway, above] {$H_{Q,k,i}$};
\node[draw, fill=red, star,star points=7,star point ratio=0.8, inner sep=5] at (2.45,0+\ZB) {};
\node[] at (3.2,0+\ZB) {tag $i$};
\end{tikzpicture}
\caption{The matrices associated with the model. For instance, $A_{k,\ell}$ corresponds to the free space loss \emph{from} antennas $k$ \emph{to} antenna $\ell$.}
\label{fig:sys_mat}
\end{figure}

Applying the same rotation to any two tags does not change the distance between them, so we can drop $Q$ from the subscript of $B_Q$ and use $B$ instead (see Figure \ref{fig:sys_mat}). Now, let $\mathbf{s}_t$ and $\mathbf{s}_t'$ be the vectors of the transmitted signals at the antennas and tags respectively at time $t$, and $\mathbf{\Psi}_t$ and $\mathbf{\Psi}_t'$ be the vectors of the received signals at the antennas and tags respectively at time $t$. The relationship between these vectors is given by:
\begin{equation}
\label{eq:signal_rel}
\begin{bmatrix}
\mathbf{\Psi}_t \\
\mathbf{\Psi}_t'
\end{bmatrix} = \begin{bmatrix}
A & H_Q \\
H_Q^T & B
\end{bmatrix}\begin{bmatrix}
\mathbf{s}_t \\
\mathbf{s}_t'
\end{bmatrix}.
\end{equation}

As mentioned before, each backscatter tag can be configured by setting its state to a value $i\in \{0,1\}$, which in turn determines its reflectivity $r_i\in \mathbb{C}$. Hence, the reflectivities can expressed in terms of the states assigned to the tags. Letting $s_{t,n}$ be the state of the $n^{\mathrm{th}}$ tag at time $t$, $\mathbf{c}_t = (s_{t,1},\ldots,s_{t,N})$ be the \emph{codeword} at time $t$ (a codeword is a vector of states), and $r(\cdot)$ be the mapping from codewords to reflectivity vectors, we define the matrix of reflectivities at time $t$ as: 
\begin{align}
\label{reflectivities}
 R_t &= \textrm{diag}(r(\mathbf{c}_t)).
\end{align}

Using the fact that the tags reflect the incident signals, the transmitted signal at the tags is given by $\mathbf{s}_t' = R_t\mathbf{\Psi}_t'$. Substituting the value of $s_t'$ in (\ref{eq:signal_rel}), we obtain the following equations: 
\begin{align}
    \mathbf{\Psi}_t &= A\mathbf{s}_t + H_QR_t\mathbf{\Psi}_t' \\
    \mathbf{\Psi}_t'&= H_Q^T\mathbf{s}_t + BR_t\mathbf{\Psi}_t'
\end{align}

Solving the above equations, we obtain that the received signal is given by $\mathbf{\Psi}_t = As_t + H_QR_t(I - BR_t)^{-1}H_Q^T\mathbf{s}_t$. Since the first term in the previous expression does not depend on $Q$, we can subtract it at the receiver\footnote{We are eliminating full-duplex self-interference, \emph{e.g.,} see \cite{Sabharwal:2014}.} yielding a modified noiseless signal $\mathbf{f}(Q;\mathbf{c}_t)\in\mathbb{C}^K$ given by: 
\begin{equation}
\label{eq:static_output}
\mathbf{f}(Q;\mathbf{c}_t) =  H_QR_t(I - BR_t)^{-1}H_Q^T\mathbf{s}_t.
\end{equation}
\subsection{Codes}
The expression in (\ref{eq:static_output}) is the \emph{noiseless} channel output when using a codeword $\mathbf{c}_t$. Based on numerical simulations, we have observed that a single codeword often only allows us to detect some orientations, and not others. Specifically, all tag arrays we came across had the following property: for every codeword $\mathbf{c}\in \{0,1\}^N$, there exists some $Q,Q'$ such that $Q\neq Q'$ and $\mathbf{f}(Q;\mathbf{c}) \approx \mathbf{f}(Q';\mathbf{c})$ (see Figure \ref{fig:tetra_array} in Section \ref{sec:numerics} for such an array). In other words, for every codeword, there is a pair of orientations that the codeword makes almost indistinguishable based on the received signal. 

Therefore, using different codewords over time is a natural extension to the model as a new codeword could fill in the gaps left by the ones used previously. Based on this intuition, we define a \emph{code} $C = [\mathbf{c}_1 \cdots \mathbf{c}_T]\in\{0,1\}^{N\times T}$ to be a binary matrix where each column corresponds to the codeword at a specific time $t$ (See Figures \ref{fig:time_code},\ref{fig:code_matrix}). Given the code $C$, we denote the \emph{concatenated} channel output as (using $F$ instead of $f$):
\begin{align}
\label{eq:concatenated_output}
F(Q;C) &= \begin{bmatrix}
\mathbf{f}(Q;\mathbf{c}_1) & 
\cdots & 
\mathbf{f}(Q;\mathbf{c}_T)
\end{bmatrix} \in \mathbb{C}^{K\times T}.
\end{align}
\subsection{Performance Measures}
\label{sec:perf_criteria}
We consider estimation over finite uniform subset of orientations $\mathcal{Q}\subset \mathrm{SO}(3)$, and use the loss function $\theta(Q,Q')$, equal to the Frobenius norm between the two rotation matrices i.e., 
\begin{equation}
\label{eq:rotation_distance}
    \theta(Q,Q') = \left\lVert Q - Q'\right\rVert_F.
\end{equation}
Letting $C$ be any code, and $W$ be complex Gaussian noise, we define the noisy received signal as:
\begin{equation}
\label{eq:noisy_output}
Y=F(Q;C) + W.
\end{equation}
Hence, the \emph{unnormalized} average error of coding matrix $C$ is given by:
\begin{equation}
\label{eq:avg_error}
\mathcal{L}(C) = \sum_{Q\in \mathcal{Q}} \mathbb{E}_{W} \left[\theta(\widehat{Q}(Y), Q)\right],
\end{equation}
where $\widehat{Q}(Y)$ is the minimum distance decoder to the grid of orientations $\mathcal{Q}$ i.e., the MMSE estimator of $Q$ given $Y$. Note if the orientation is considered a random variable, we implicitly assume a uniform prior over the orientations in $\mathcal{Q}$. Now, a code that minimizes the average error could still provide poor performance for a \emph{particular} orientation. For this reason, we study the worst-case error as it delivers guarantees on the performance across all possible orientations. We define worst-case error (over all possible orientations in $\mathcal{Q}$) as: 
\begin{equation}
    \label{eq:worst_error}
    \mathcal{M}(C) = \max_{Q\in \mathcal{Q}} \mathbb{E}_W \left[\theta(\widehat{Q}(Y), Q)\right].
\end{equation}
We believe that both quantities are of interest for designing an orientation estimation system and hence, we suggest criteria for minimizing either depending on the application's particular requirements.
\begin{figure}[t!]
\centering
\vspace{0.2cm}
\input{table}
\vspace{0.25cm}
\caption{An illustration of coding over time. Different colors denote different states (reflectivities) of the tag, and different patterns of the states denote different codewords.}
\label{fig:time_code}
\end{figure}
\begin{figure}[t!]
\vspace{-0.6cm}
\centering
\pgfmathsetmacro{\myscale}{0.8}
\pgfkeys{tikz/mymatrixenv/.style={decoration={brace},every left delimiter/.style={xshift=8pt},every right delimiter/.style={xshift=-8pt}}}
\pgfkeys{tikz/mymatrix/.style={matrix of math nodes,nodes in empty cells,
left delimiter={[},right delimiter={]},inner sep=1pt,outer sep=1.5pt,
column sep=8pt,row sep=8pt,nodes={minimum width=20pt,minimum height=10pt,
anchor=center,inner sep=0pt,outer sep=0pt,scale=\myscale,transform shape}}}
\pgfkeys{tikz/mymatrixbrace/.style={decorate,thick}}

\newcommand*\mymatrixbraceright[4][m]{
    \draw[mymatrixbrace] (#1.west|-#1-#3-1.south west) -- node[left=2pt] {#4} (#1.west|-#1-#2-1.north west);
}
\newcommand*\mymatrixbraceleft[4][m]{
    \draw[mymatrixbrace] (#1.east|-#1-#2-1.north east) -- node[right=2pt] {#4} (#1.east|-#1-#2-1.south east);
}
\newcommand*\mymatrixbracetop[4][m]{
    \draw[mymatrixbrace] (#1.north-|#1-1-#2.north west) -- node[above=2pt] {#4} (#1.north-|#1-1-#3.north east);
}
\newcommand*\mymatrixbracebottom[4][m]{
    \draw[mymatrixbrace] (#1.south-|#1-1-#2.north east) -- node[below=2pt] {#4} (#1.south-|#1-1-#3.north west);
}

\tikzset{greenish/.style={
    fill=green!50!lime!60,draw opacity=0.4,
    draw=green!50!lime!60,fill opacity=0.1,
  },
  cyanish/.style={
    fill=cyan!90!blue!60, draw opacity=0.4,
    draw=blue!70!cyan!30,fill opacity=0.1,
  },
  orangeish/.style={
    fill=orange!90, draw opacity=0.8,
    draw=orange!90, fill opacity=0.3,
  },
  brownish/.style={
    fill=brown!70!orange!40, draw opacity=0.4,
    draw=brown, fill opacity=0.3,
  },
  purpleish/.style={
    fill=violet!90!pink!20, draw opacity=0.5,
    draw=violet, fill opacity=0.3,    
  },
  redish/.style={
    fill=red!90, draw opacity=0.5,
    draw=red, fill opacity=0.3,    
  }}
\centering
\vspace{-0.25cm}
\hspace{1.4cm}
\input{matrix.tikz}
\caption{The associated matrix to the coding over time example in Figure $4$.}
\label{fig:code_matrix}
\end{figure}
\section{Main Results}
\label{sec:results}
Our goal is to find the best code for the estimation task. However, $\mathcal{L}(C)$ in (\ref{eq:avg_error}) and $\mathcal{M}(C)$ in (\ref{eq:worst_error}) are intractable. One route we could take is to investigate \emph{tractable} upper or lower bounds on the errors, and minimize those bounds as a proxy for minimizing the errors themselves.
\subsection{Design Criterion For The Average Error}
\label{sec:average_code}
As a proxy for the average error $\mathcal{L}(C)$, we minimize an upper bound to it. We obtain the upper bound on the average error by replacing the probability of misestimating the actual orientation of the object by a tractable quantity. The proof of the following result is given in Section \ref{sec:proof_upper}.
\begin{theorem}
\label{thm:upper}
Assuming complex white Gaussian noise, the average error $\mathcal{L}(C)$ given in (\ref{eq:avg_error}) is upper bounded by:
\begin{align}
\label{eq:upper_bound}
\mathcal{U}(C) = \sum_{Q,Q'\in \mathcal{Q}} \mathrm{erfc}\left(\frac{\left\lVert F(Q;C) - F(Q';C) \right\rVert_F}{2\sqrt{2}\sigma}\right) \theta(Q, Q').
\end{align}
where $\theta(Q,Q')$ is given in (\ref{eq:rotation_distance}), $\mathrm{erfc}$ is the complement of the Gaussian error function, and $\sigma$ is the standard deviation of the noise.
\end{theorem}
Hence, the design criteria for minimizing the average error is simply given by the optimization problem $\min_C \mathcal{U}(C)$.
We can infer the following intuition from $\mathcal{U}(C)$ in (\ref{eq:upper_bound}): good codes should map orientations that are far apart (in terms of $\theta$) to channel outputs that are also far apart. Moreover, since $\mathrm{erfc}(x)$ is tightly upper bounded by $\mathrm{exp}(-x^2)$, $\mathcal{U}(C)$ implies that the distance between channel outputs  $\left\lVert F(Q;C) - F(Q';C) \right\rVert_F$ has an inverse exponential relationship with the average error.
\subsection{Codes For the Worst-Case Error}
Whereas we optimize an upper bound for the average error in Section \ref{sec:average_code}, we obtain a lower bound on the worst-case error $\mathcal{M}(C)$ using Le Cam's method\cite{shiryayev1989}, and minimize the bound to obtain a code for the worst-case performance. The proof of the following theorem is provided in Section \ref{sec:proof_worst}. 
\begin{theorem}
\label{thm:worst}
Assuming complex white Gaussian noise, the worst-case error $\mathcal{M}(C)$ given in (\ref{eq:worst_error}) is lower bounded by: 
\begin{multline}
    \label{eq:lower_bound}
    \mathcal{V}(C) = \max_{Q,Q'\in \mathcal{Q}}\mathrm{exp}\left({-\frac{\left\lVert F(Q;C) - F(Q';C)\right\rVert_F^2}{2\sigma^2}}\right)\\
    \frac{\theta(Q, Q')}{4},
\end{multline}
where $\theta(Q,Q')$ is given in (\ref{eq:rotation_distance}).
\end{theorem}
Hence, the design criteria for minimizing the worst-case error is given by the optimization problem $\min_C \mathcal{V}(C)$. $\mathcal{V}(C)$ corroborates the observations from Theorem \ref{thm:upper}, and relates the worst-case error to the model only through the distances between channel outputs. $\mathcal{V}(C)$ again implies that the worst-case error has an inverse exponential relationship with $\left\lVert F(Q;C) - F(Q';C) \right\rVert_F$.
\subsection{Minimax Bound}
Unlike Theorem \ref{thm:worst} which suggests a design criteria for the worst-case error using Le Cam's method, the following theorem uses the same tools to quantify the worst-error error decay with parameters including the number of antennas $K$, the number of tags $N$ through the Frobenius norm of $X^{\mathrm{tag}}$, and the number of samples through the variance $\sigma^2$ (as the effective variance of the noise is $\sigma^2/n$ when provided with $n$ samples). It also reflects the effect of code design on the error. 
\begin{theorem}[Minimax Bound] 
\label{thm:Lecam}
The worst-case error is bounded as:  
\begin{align}
\label{eq:LecCamBound}
\mathcal{M}(C) \geq \frac{32\pi^2\lambda^2\sigma^2 D^4}{27K^2 \left\lVert X^{\mathrm{tag}}\right\rVert_F^2 \sum_{t=1}^T \lVert \widetilde{B}_t\rVert_F^2\left\lVert r(\mathbf{c}_t)\right\rVert^2}
\end{align}
where $\widetilde{B}_t= (I-BR_t)^{-1}$ and $D$ is an approximate range between the tagged object and the antennas.
\end{theorem}
The bound indicates that placing the tags far away from the antennas leads to a larger error. This is clear as the farther away the tags, the weaker the signal we receive due to attenuation. Moreover, it implies an inverse quadratic relationship between the number of antennas and the error. This is again logical because we would be able to receive more power from the reflected signal with more antennas. The bound also reflects the effect of code design through the term $\Vert \tilde{B}_t \Vert_{F}^{2}\left\lVert r(\mathbf{c}_t)\right\rVert^2$. For instance, if tags can have each have one of two states with corresponding reflectivities $0$ and $1$ (on or off), then the bound favors most tags to be "on" through the term $\left\lVert r(\mathbf{c}_t)\right\rVert^2$.
\subsection{Structure of the optimal codes}
Finding the codes that minimize $\mathcal{U}(C)$ and $\mathcal{V}(C)$ is a combinatorial optimization problem that requires an exhaustive search. Performing an exhaustive search has a running time of $\mathcal{O}(2^{NT})$ if wish to look for the best coding matrix of size $T$. Hence, for our purposes, finding the best code using brute-force is intractable except for cases when the number of tags $N$, and the size of the coding matrix $T$, are very small.

Therefore, we introduce the idea of code proportions which is essential in reducing the search space. We show that our performance measures depend only on the code through particular proportions of each codeword (see Theorem \ref{thm:dependence}), therefore reducing the search space to only those proportions. We begin by carefully defining this idea. Let $C = [\mathbf{c}_1 \cdots \mathbf{c}_T]$ be any code, then we define the proportion of code configuration $\mathbf{c}\in \{0,1\}^N$ in $C$ as
\begin{align}
    \pi_\mathbf{c} = \frac{1}{T}\sum_{t=1}^{T} 1_{\mathbf{c} = \mathbf{c}_t},
\end{align}
where $1_{\mathbf{c}=\mathbf{c}_t}$ is the indicator of the event $\{\mathbf{c} = \mathbf{c}_t\}$. In other words, $\pi_\mathbf{c}$ is the number of occurrences of $\mathbf{c}$, normalized by $T$, the size of the code. Moreover, let $\mathbf{c}^{(1)},\ldots, \mathbf{c}^{\left(M\right)}$ be an enumeration of the codewords in $\{0,1\}^N$, then we define the vector of code proportions as $\mathbf{\pi} = \begin{bmatrix} \pi_{\mathbf{c}^{(1)}}, \cdots, \pi_{\mathbf{c}^{\left(M\right)}}\end{bmatrix}^T$. It is straightforward to see that if two coding matrices share the same size $T$, and the same proportions vector $\mathbf{\pi}$, then they are permutations of one another. In what follows, instead of directly using the coding matrix $C$, we search for the optimal coding scheme through the proportions vector $\pi$. 

The following theorem allows us to analyze the effect of $C$ on the average and worst-case errors through the proportions vector $\mathbf{\pi}$. Moreover, we show that the derived bounds in (\ref{eq:upper_bound}) and $(\ref{eq:lower_bound})$ can be written in terms of $\pi$. The proofs of the following results is given in Section \ref{sec:proof_dependence}.
\begin{theorem}
\label{thm:dependence}
Assuming independent and identically distributed noise, $\mathcal{L}(C)$ and $\mathcal{M}(C)$ only depend on coding matrix $C$ through the proportions vector $\mathbf{\pi}$.
\end{theorem}
\begin{lemma}
\label{lem:rewrite}
If we fix $T$, the size of the code $C$, then we can write the minimization of $\mathcal{U}(C)$ and $\mathcal{V}(C)$ in term of $\pi$ respectively as:
\begin{align}
 \label{eq:upper_bound_rewrite}
    \mathcal{U}(\pi) &= \sum_{Q,Q'\in \mathcal{Q}} \mathrm{erfc}\left(\sqrt{\frac{T\pi \mathbf{g}(Q,Q')}{8\sigma^2}}\right) \theta(Q,Q'),\\
     \label{eq:lower_bound_rewrite}
    \mathcal{V}(\pi) &=\max_{Q,Q'\in \mathcal{Q}} \mathrm{exp}\left(-\frac{T\mathbf{\pi}^T\mathbf{g}(Q,Q')}{2\sigma^2}\right)\theta(Q,Q'),
\end{align}
where $\mathbf{g}(Q,Q')\in \mathbb{R}^M$ is given by $g(Q,Q')_i =\left\lVert f(Q;\mathbf{c}^{(i)}) - f(Q';\mathbf{c}^{(i)})\right\rVert$. 
\end{lemma}
Theorem \ref{thm:dependence} and Lemma \ref{lem:rewrite} allow us to write our design criteria as optimization problems with respect to the proportions vector $\pi$. In other words, the design criteria for the average and worst-case errors are equivalent to $\min_{\pi} \mathcal{U}(\pi)$ and $\min_{\pi}\mathcal{V}(\pi)$, respectively. In (\ref{eq:upper_bound_rewrite}) and (\ref{eq:lower_bound_rewrite}), $\mathbf{g}(Q,Q')$ represents the link between the design criteria channel model. This implies that our design criteria will depend on the channel model only through the distances between channel outputs for different values of the orientation.
\section{Numerical Simulations}
\label{sec:numerics}

\subsection{Simulation Setup}
\begin{figure}[h]
    \centering
    \includegraphics[scale=1]{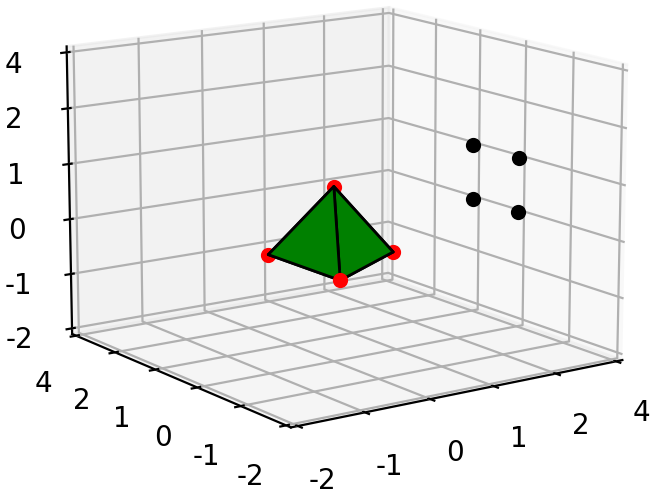}
    \caption{An example of a simulation setup. The tags (given by the red dots) are in a tetrahedron configuration, while the antennas (given by the black dots) are in a square configuration.}
    \label{fig:tetra_array}
\end{figure}

We use $N=4$ backscatter tags that we place randomly within a sphere of radius $0.25$m. The 4 tags can \emph{each} have two states, with reflection coefficients $-0.5$ and $0.5$, encoded as $0$ and $1$, respectively. We arrange $K=4$ full-duplex antennas in a $1\mathrm{m}\times1\mathrm{m}$ square on a plane $4$m away from the center of the object (see Figure \ref{fig:tetra_array} for a sample arrangement). Each antenna emits an identical signal $s_k = 1$ for all $k=1,\ldots,K$. We use a wavelength $\lambda = 0.005$m, and generate a set of orientations $\mathcal{Q}$ with $\left\lvert \mathcal{Q}\right\rvert=4000$ by uniformly sampling the ranges of the euler angles  and computing the rotation matrices that correspond to the angles. 

We define SNR as \emph{signal-to-noise ratio}, the received signal strength at each antenna, divided by the noise power. That is, given the signal-noise ratio SNR, the noise variance is given by $K/\mathrm{SNR}$, where $K$ is the number of antennas. We test each value of $\mathrm{SNR}=0,1,\ldots,10$ dB, with $500$ trials for each of the $4000$ "ground truth" orientations in $\mathcal{Q}$. As a measure of distance, we use the polar and azimuthal angles in the polar coordinate system $(r,\theta,\varphi)$. It is necessary to define a unique set of spherical coordinates for each point, so we restrict the ranges of the polar angle and azimuthal angle to $[0, \pi)$ and  $[0, 2\pi)$, respectively. All things considered, the measure of distance we use in our trials is given by: 
\begin{equation}
    \ell \left[(\theta,\varphi), (\hat{\theta}, \hat{\varphi})\right] = (\theta-\hat{\theta})^2 + (\varphi - \hat{\varphi})^2,
\end{equation}
where $(\theta,\varphi)$ are the ground truth angles, and $(\hat{\theta},\hat{\varphi})$ are the estimated angles. To obtain an estimate $\widehat{\mathcal{L}}(C)$ of the average-case error $\mathcal{L}(C)$, we average the value of the error over the $500\times 4000 =2,000,000$ trials. The performance of a code varies across different orientations. Hence, unique to average error trials, we have included the variance of each coding method at an SNR value of $10\mathrm{dB}$. The variance indicates how much the performance varies from the total average across the different tested orientations. On the other hand, to obtain an estimate $\widehat{\mathcal{M}}(C)$ of the worst-case error $\mathcal{C}$, we average the $500$ trials for each of the considered orientations, and then take the maximum over all the orientations in $\mathcal{Q}$. In these simulation trials, we use codes of size $T=24$, and compare the performance of the following methods: 
\begin{itemize}
    \item REP\_OPT: The best repetition code (all $T=24$ codewords used are the same). This method is equivalent to the design criteria suggested in \cite{CRSD-ISIT20}.
    \item ORTHOGONAL: The code where each of the vectors in the standard basis $\{e_1,\ldots,e_4\}$ is used $6$ times. Note that $e_i$ is the vector components whose are all zero, except for a $1$ in the $i^{th}$ component. 
    \item AVERAGE\_DESIGN: The method we suggested to minimize the average error. This is the coding scheme that minimizes the upper bound of the average error i.e. $\pi^*=\arg\min_{\pi} \mathcal{U}(\pi)$.
    \item MINIMAX\_DESIGN: The method we suggested to minimize the worst-case error. This is the coding scheme that minimizes the lower bound of the worst-case error i.e. $\pi^*=\arg\min_{\pi} \mathcal{V}(\pi)$.
\end{itemize}
\subsection{The Average Error}
\begin{figure}[t!]
    \centering
    \begin{minipage}{0.5\textwidth}
        \centering
        \includegraphics[width=0.95\textwidth]{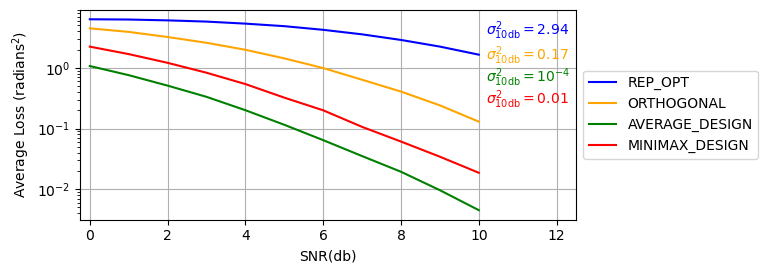} 
    \end{minipage}
    \caption{The average squared error versus SNR for the tag array in Figure \ref{fig:tetra_array}. At an $\mathrm{SNR}$ value of $10\mathrm{dB}$, the designed code for the average error is nearly $70\times$ more accurate than the channel-agnostic orthogonal code.}
    \label{fig:average_tetra}
\end{figure}

\begin{figure}[t!]
    \centering
    \hspace{-1.35cm}
    \begin{minipage}{0.5\textwidth}
        \centering
        \includegraphics[width=0.8\textwidth]{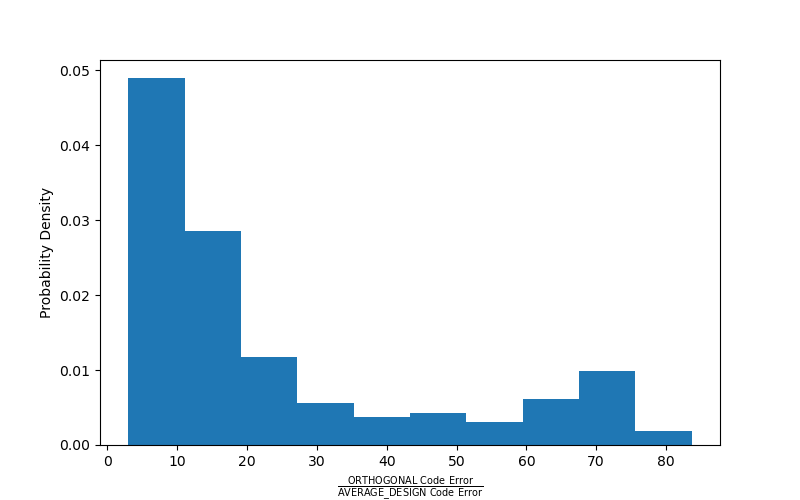} 
    \end{minipage}
    \caption{Probability density histogram of the average error ratios when using the orthogonal code and the designed code at an SNR of $10\mathrm{dB}$.}
    \label{fig:average_histogram}
\end{figure}

The results shown in Figures \ref{fig:average_tetra} and \ref{fig:average_histogram} suggest that channel knowledge at the receiver provides significant benefits over orthogonal codes. For the average error, the code we designed for the average error lead to the best performance out of the considered methods, with the code produced by the method we suggested for the worst-case error trailing behind as a close second. In particular, for the tetrahedron tag configuration shown in Figure \ref{fig:tetra_array}, the average error when using the code produced by the optimization problem $\min_\pi \mathcal{U}(\pi)$ is $70\times$ lower than the error when using orthogonal code. Furthermore, the performance obtained when using the best static code is consistently the worst out of the considered methods. This corroborates the intuition that lead us to introduce the idea of different codes over time. 

In order to test the effectiveness of the designed criteria on tag arrays different from the one shown in Figure \ref{fig:tetra_array}, we randomize the positions of the tags (within a sphere of radius 0.25m), compute the designed code, and calculate the average error ratio obtained when using orthogonal codes and the designed code. We perform these steps for a total of 200 times, and produce the density histogram shown in Figure \ref{fig:average_histogram}. In almost all of the 200 trials, the designed code offered more than a $5\times$ improvement over the channel oblivious orthogonal code, and in almost half of the trials, the designed code offered more than a $10\times$ improvement. 
\subsection{The Worst-Case Error}
\begin{figure}[t!]
    \centering
    \begin{minipage}{0.5\textwidth}
        \centering
        \includegraphics[width=0.95\textwidth]{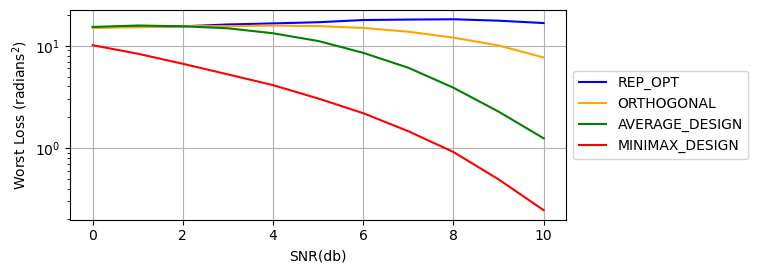} 
    \end{minipage}
    \caption{The worst-case error versus SNR for the tag array in Figure \ref{fig:tetra_array}. At an $\mathrm{SNR}$ value of $10\mathrm{dB}$, the designed code for the worst-case error is nearly $100\times$ more accurate than the orthogonal code.}
    \label{fig:worst_tetra}
\end{figure}

Moreover, the results shown in Figure \ref{fig:worst_tetra} suggest that channel knowledge for code design once again provides large benefits. For the tetrahedron array of Figure \ref{fig:tetra_array},  the worst-case error produced by the optimization problem $\min_\pi \mathcal{V}(\pi)$ is $100\times$ lower than the error produced when using orthogonal codes.

\begin{figure}[t!]
    \centering
    \vspace{0.1 cm}
    \hspace{-1.35cm}
    \begin{minipage}{0.5\textwidth}
        \centering
        \includegraphics[width=0.8\textwidth]{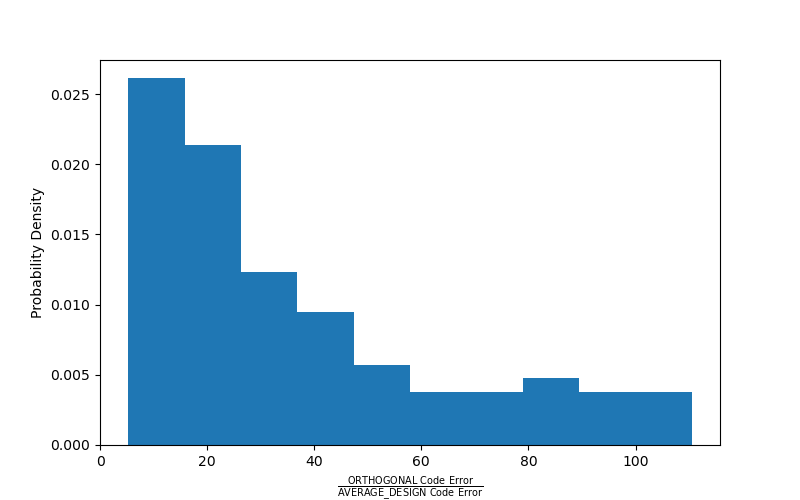} 
    \end{minipage}
    \caption{Probability density histogram of the worst error ratios when using the orthogonal code and the designed code at an SNR of $10\mathrm{dB}$.}
    \label{fig:worst_histogram}
\end{figure}

As we did before, we randomize the positions of tags, and compute the worst error ratio obtain when using orthogonal codes and the code that minimizes (\ref{eq:lower_bound_rewrite}). We repeat these steps for a total of 150 times, and produce the density histogram in Figure $\ref{fig:worst_histogram}$. In almost all of the 150 trials, the designed code offered a $5\times$ advantage over orthogonal codes, and in almost half of the trials, the designed code offered more than a $20\times$ improvement. 
\subsection{Multipath}
\begin{figure*}[t!]
    \centering
    \begin{minipage}{0.5\textwidth}
        \centering
        \includegraphics[width=0.9\textwidth]{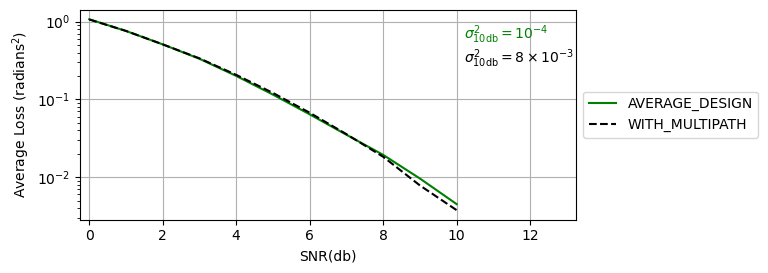} 
    \end{minipage}\hfill
    \begin{minipage}{0.5\textwidth}
        \centering
        \includegraphics[width=0.9\textwidth]{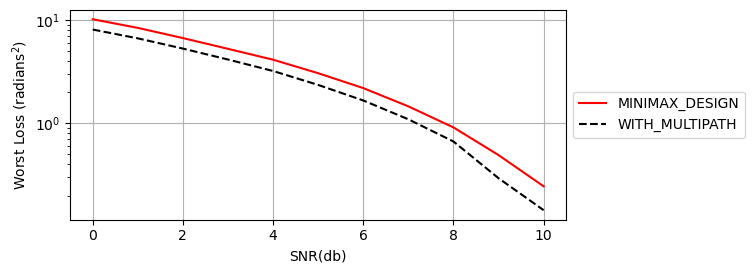} 
    \end{minipage}
    \caption{The average (left) and worst-case (right) errors for the designed codes in the cases of line-of-site only and multipath.}
    \label{fig:multipath}
\end{figure*}

Suppose we now add $M$ reflectors with positions $\mathbf{x}_1^{\textrm{ref}},\ldots, \mathbf{x}_M^\textrm{ref}\in\mathbb{R}^{3}$ to our system. We treat reflectors as virtual sources, and use the following loss formula for the path that begins at $x_1$, is reflected at $x^{\mathrm{ref}}$, and ends at $x_2$:
\begin{equation}
\gamma(x_1,x^{\textrm{ref}},x_2) = \eta(x_1,x^{\textrm{ref}})\eta(x^{\textrm{ref}},x_2).
\end{equation}
Let $D_{m,Q}$ be the $K\times N$ matrix of the multipath only channel responses between the tags and antennas in the presence of $x^{\textrm{ref}}_m$, i.e.
\begin{align}
(D_{Q,m})_{k,n} &= \gamma(x_k^{\textrm{ant}}, x_m^{\textrm{ref}},Qx_n^{\textrm{tag}}).
\end{align}
Now, the matrices involved in our channel model remain precisely the same, except for $H_Q$ which is now replaced by $E_Q = H_Q + D_Q$. The channel model is thus extended to: 
\begin{equation}
f_m(Q;\mathbf{c}_t) =  E_QR_t(I - BR_t)^{-1}E_Q^T\mathbf{s}_t.
\end{equation}
In our trials, we use 10 multipath components placed 1m away from the object. Based on observations, and the results of Figure \ref{fig:multipath}, multipath does little in the way of diversity gain. However, given a fixed transmit power, multipath does offer an SNR gain at the receiver. 
\subsection{Robustness}
\begin{figure*}[t!]
    \centering
     \begin{minipage}{0.5\textwidth}
        \centering
        \includegraphics[width=0.9\textwidth]{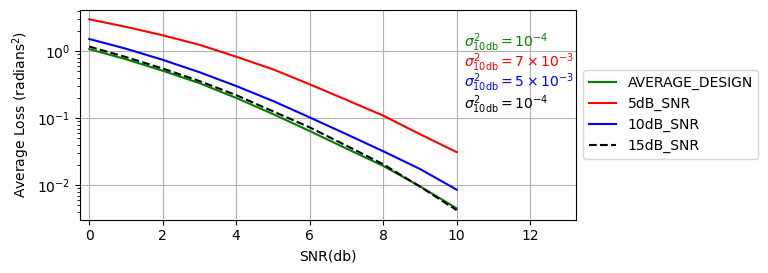} 
    \end{minipage}\hfill
    \begin{minipage}{0.5\textwidth}
        \centering
        \includegraphics[width=0.9\textwidth]{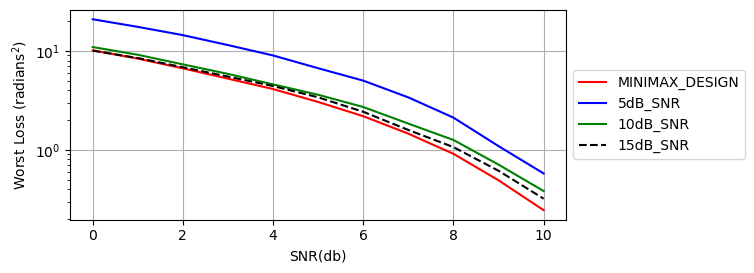} 
    \end{minipage}

    \caption{The average (left) and worst-case error (right) of the designed codes for different levels of channel estimation errors.}
    \label{fig:robust}
\end{figure*}

Computing the codes schemes suggested by our design criteria requires channel knowledge. However, since we are not likely to have perfect channel knowledge, we test the performance of our design criteria when computed using imperfect channel estimation. We introduce channel estimation errors by adding noise to the channel outputs in different orientations. Specifically, instead of using $F(Q;C)$ to compute the suggested coding scheme, we use  
\begin{equation}
    F'(Q;C) = F(Q;C) + W_Q,
\end{equation}
where $W_Q$ is white Gaussian noise, and $W_Q$ is independent of $W_{Q'}$ for every $Q'\neq Q$. We tested the following values of the $\mathrm{SNR}= 5,10,15\mathrm{dB}$, with $50$ trials each, and then averaged the performance. In particular, the results of Figure \ref{fig:robust} indicate that our designed codes retain nearly all even when computed with channel estimation errors greater than or equal to $10$dB. The figures also indicate that the design criteria maintain considerable improvement over channel oblivious codes when computed with channel estimation errors less than $10\mathrm{dB}$.
\section{Proofs}
\label{sec:proofs}
\subsection{Proof of Theorem \ref{thm:upper}}
\label{sec:proof_upper}
\noindent Let $Y = F(Q,C) + W$ be as in \ref{eq:noisy_output}. $\{\widehat{Q}(Y) = Q'\}$ is the event that our estimator outputs a different orientation $Q'$ when the object is in orientation $Q$. We can rewrite the average error in terms of events of this form.
\begin{align}
\mathcal{L}(C) &= \sum_{Q\in \mathcal{Q}} \mathbb{E}_W \left[\theta(\widehat{Q}(Y), Q)\right],\\ \label{eq:different_form}
&= \sum_{Q\in\mathcal{Q}}\sum_{Q'\neq Q}\mathbb{P}\left(\widehat{Q}(Y) = Q'\right)\theta(Q, Q').
\end{align}
The probability of the event $\{\widehat{Q}(Y) = Q'\}$ is a difficult quantity to compute outside of cases when $\mathcal{Q}$ is very small. Therefore, we wish to upper bound it with another probability that is easier to compute. Recall that 
\begin{align}
    \widehat{Q}(Y) =\arg\min_{Q\in \mathcal{Q}} \left\lVert Y - F(Q,C)\right\rVert_F.
\end{align}
Given the functional form of the estimator, a necessary but insufficient condition for our estimator to output $Q'$ is for the received signal $Y$ to satisfy $\left\lVert Y - F(Q;C) \right\rVert \geq  \left\lVert Y - F(Q';C) \right\rVert$. Given this relationship between the events, we have that:  
\begin{align}
\label{eq:upperboundevent}
\big\{ \widehat{Q}(Y) = Q' \big\} &\subseteq \big\{ \left\lVert Y - F(Q;C) \right\rVert_F \geq  \left\lVert Y - F(Q';C) \right\rVert_F \big\}.
\end{align}
The latter is event that our \emph{noisy} channel output is closer to the \emph{noiseless} channel output corresponding to $Q'$ than the \emph{noiseless} channel output corresponding to $Q$. Combining (\ref{eq:different_form}) and (\ref{eq:upperboundevent}) with the monotonicity of the probability function ($\mathbb{P}(A) \leq \mathbb{P}(B)$ for all events $A\subseteq B$), we obtain an upper bound on the average error. 
\begin{align}
\mathcal{L}(C) = \sum_{Q\in\mathcal{Q}}\sum_{Q'\neq Q}&\mathbb{P}\left(\widehat{Q}(Y) = Q'\right)\theta(Q, Q'), \\
\leq\sum_{Q\in\mathcal{Q}}\sum_{Q'\neq Q}&\mathbb{P}\big(\left\lVert Y - F(Q;C) \right\rVert_F \geq  \left\lVert Y - F(Q';C) \right\rVert_F \big)\nonumber \\
\label{eq:upperbound2}
&\theta(Q, Q').
\end{align}
In other words, we use the standard pairwise error probability to develop an upper bound to the average error. Combining (\ref{eq:upperbound2}) with the following standard lemma \cite{10.5555/2765665}, which provides the form for the probability of the event in the right-hand-side of (\ref{eq:upperboundevent}), we obtain the desired result.
\begin{lemma}
\label{lem:erfc}
Let $Y \sim \mathbb{C}\mathcal{N}(\mathbf{0}, \sigma^2I_n)$, and let $a\in \mathbb{C}^n$, then 
\begin{align}
    \mathbb{P}\left( \left\lVert Y \right\rVert \geq \left\lVert Y - a\right\rVert\right) = \frac{1}{2} \mathrm{erfc}\left(\frac{\left\lVert a\right\rVert}{2\sqrt{2}\sigma}\right),
\end{align}
where $\mathrm{erfc}$ is the complement of the Gauss error function. 
\end{lemma}
\subsection{Proof of Theorem \ref{thm:worst}}
\label{sec:proof_worst}
\noindent Let $C\in[m]^{N\times T}$ be a coding matrix, and let $\mathcal{P}$ be the family of distributions given by
\begin{align}
\label{eq:prob_family}
\mathcal{P}= \{\mathbb{C}\mathcal{N}(F(Q;C), \sigma^2I); Q\in \mathcal{Q}\}.
\end{align}
The probability distributions in $\mathcal{P}$ can be parameterized by their orientation. Let $P,P'\in\mathcal{P}$ be distributions with $Q$ and $Q'$ as their corresponding orientations, then Le Cam's method\cite{shiryayev1989} asserts that: 
\begin{align}
\mathcal{M}(C) &\geq \frac{\theta(Q,Q')}{2} \left( 1 - \mathrm{TV}(P,P')\right),
\end{align}
where $\mathrm{TV}(P,P')$ is the total variation distance\cite{10.5555/1146355}. The Kullback-Leibler divergence\cite{4557285} is a different and widely used measure of discrepancy between two probability distributions. It has many desirable properties, and it's an upper bound for many other popular discrepancy measures. Through Pinsker's inequality\cite{10.5555/1146355}, one can upper bound the total-variation distance through the Kullback-Leibler divergence. The total-variation distance between two probability distributions $P$ and $P'$, the Kullback-Leibler divergence, and Pinsker's inequality are respectively given by: 
\begin{align}
   &\mathrm{TV}(P,P') = \int_{\mathbb{C}^{KT}} \left\lvert P(x) - P'(x)\right\rvert\mathrm{d}x. \\
    &\mathrm{KL}(P||P') = \int_{\mathbb{C}^{KT}} P(x) \log\frac{P(x)}{P'(x)}\mathrm{d}x. \\
    &\left\lVert P - P'\right\rVert_{\mathrm{TV}} \leq \sqrt{\frac{1}{2}\mathrm{KL}(P||P')}.
\end{align}
Therefore, 
\begin{align}
    &\mathcal{M}(C) \\
    &\geq \frac{\theta(Q,Q')}{2} \left( 1 - \mathrm{TV}(P,P')\right), \\
    \label{eq:pinsker_step}
    &\geq \frac{\theta(Q,Q')}{2} \left( 1 - \sqrt{\frac{1}{2}\mathrm{KL}(P||P')}\right),\\
    \label{eq:minus_step}
    &\geq \frac{\theta(Q,Q')}{4} \mathrm{exp}\left(-\mathrm{KL}(P||P')\right), \\
    \label{eq:subst_step}
    &= \frac{\theta(Q,Q')}{4}\mathrm{exp}\left(-\frac{\left\lVert F(Q;C) - F(Q';C)\right\rVert_F^2}{2\sigma^2}\right).
\end{align}
(\ref{eq:pinsker_step}) follows from Pinsker's inequality, and (\ref{eq:minus_step}) follows from the inequality $1-x\leq e^{-x}$. For two independent Gaussians $P = \mathbb{C}\mathcal{N}(\mu, \sigma^2I)$ and $P'= \mathbb{C}\mathcal{N}(\mu', \sigma^2I)$ with the same co-variance matrix, the Kullback-Leibler divergence is given by $\left\lVert \mu - \mu' \right\rVert^2/2\sigma^2$\cite{10.5555/1146355}, and therefore, we obtain (\ref{eq:subst_step}). Since (\ref{eq:subst_step}) holds for every $P,P'\in\mathcal{P}$, we can take the maximum over every pair of distributions (equivalently, every pair of orientations $Q,Q'\in\mathcal{Q}$), which gives
\begin{multline}
    \mathcal{M}(C) \geq \max_{Q,Q'\in\mathcal{Q}}
     \mathrm{exp}\left(\frac{-\left\lVert F(Q;C) - F(Q';C)\right\rVert_F^2}{2\sigma^2}\right)\\\frac{\theta(Q,Q')}{4}.
\end{multline}
\subsection{Proof of Theorem \ref{thm:Lecam}}
\noindent Letting $\mathcal{P}$ be the same family of distributions as in (\ref{eq:prob_family}), $P$ and $P'$ distributions in $\mathcal{P}$ with $Q$ and $Q'$ as their corresponding distributions, then we can start with Equation (\ref{eq:pinsker_step}) which was obtained with Le Cam's method and Pinsker's inequality. In other words, we start with:
\begin{align}
\label{eq:2ndstep}
\mathcal{M}(C) &\geq\frac{\theta(Q,Q')}{2} \left(1-\sqrt{\frac{1}{2}\mathrm{KL}(P||P')}\right),
\end{align}
where $\mathrm{KL}(P||P')$ is the Kullback-Leibler Divergence. We obtain (\ref{eq:2ndstep}) using Pinsker's inequality\cite{10.5555/1146355}. For two independent Gaussians $P = \mathbb{C}\mathcal{N}(\mu,  \sigma^2I)$ and $P'=\mathbb{C}\mathcal{N}(\mu', \sigma^2I)$ with the same covariance matrix, $\mathrm{KL}(P,P') = \left\lVert \mu - \mu'\right\rVert_2/2\sigma^2$. Hence,
\begin{equation}
    \mathcal{M}(C) \geq \frac{\theta(Q,Q')}{2}\left(1 - \frac{1}{2\sigma}\left\lVert F(Q;C) - F(Q';C)\right\rVert_F\right).
\end{equation}
Now, let $R_t$ be the diagonal matrix of reflectivities when using code configuration $\mathbf{c}_t$, then recall
\begin{align}
f(Q;\mathbf{c}_t) &= H_{Q}R(I-BR_t)^{-1}H_{Q}^T\mathbf{s}_t, \\
&= H_{Q}(I-BR_t)^{-1}R_tH_{Q}^T\mathbf{s}_t, \\
&= H_{Q}(I-BR_t)^{-1}\textrm{diag}\left\{H_{Q}^T\mathbf{s}_t\right\}r(\mathbf{c}_t),
\end{align}
where $r(\mathbf{c}_t)$ is the column vector of reflectivities corresponding to code $\mathbf{c}_t$. We wish to bound $\left\lVert f(Q;\mathbf{c}_t) - f(Q';\mathbf{c}_t)\right\rVert$, so we use the inequality $\left\lVert A_t\mathbf{x}_t\right\rVert\leq \left\lVert A_t\right\rVert_F\left\lVert \mathbf{x}_t\right\rVert$ with $\mathbf{x}_t = r(\mathbf{c}_t)$ and
\begin{align*}
A_t = &H_{Q}(I-BR_t)^{-1}\textrm{diag}\{H_{Q}\mathbf{s}_t\} \\
    &- H_{Q'}(I-BR_t)^{-1}\textrm{diag}\{H_{Q'}\mathbf{s}_t\}.
\end{align*}
\begin{lemma}
\label{lem:ugly}
If $D$ is an approximate range between the tagged object's hinge point and the aperture, then for all $t\in \{1,\ldots,T\}$
\begin{equation}
\label{eq:frobenius_bound}
\left\lVert A_t\right\rVert_F \leq \left(\frac{K}{4\pi\lambda D^2}\right) \left\lVert X^{\mathrm{tag}}\right\rVert_F \left\lVert \widetilde{B}_t\right\rVert_F \theta(Q,Q').
\end{equation}
\end{lemma}
\noindent Letting $\delta = \sqrt{\theta(Q,Q')}$, we have that: 
\begin{align}
&\left\lVert F(Q;C) - F(Q';C)\right\rVert_F \\
&= \sqrt{\sum_{t=1}^T \left\lVert f(Q;\mathbf{c}_t) - f(Q';\mathbf{c}_t)\right\rVert } \\
\label{eq:sstep1}
&\leq \sqrt{\sum_{t=1}^T  \left\lVert A_t\right\rVert_F^2\left\lVert r(\mathbf{c}_t)\right\rVert^2}\\
\label{eq:sstep2}
&\leq \delta \sqrt{\sum_{t=1}^T  \underbrace{\left(\frac{K}{2\pi\lambda D^2} \left\lVert X^{\mathrm{tag}}\right\rVert_F \left\lVert \widetilde{B}_t\right\rVert_F \left\lVert r(\mathbf{c}_t)\right\rVert\right)^2}_{\Delta_t}},
\end{align}
where (\ref{eq:sstep1}) follows from the inequality $\left\lVert A\mathbf{x}\right\rVert \leq \left\lVert A\right\rVert_F\left\lVert \mathbf{x}\right\rVert$, and (\ref{eq:sstep2}) follows from Lemma \ref{lem:ugly}. Hence, a lower bound on the worst-case error is given by: 
\begin{equation}
\label{eq:bound_max}
    \mathcal{M}(C) \geq \frac{\delta^2}{2} \left(1 - \frac{\delta}{2\sigma}\sqrt{\sum_{t=1}^{T}\Delta_t}\right).
\end{equation}
To obtain a tightest bound, we have to find the maximum of the right-hand side with respect to $\delta$. Maximizing (\ref{eq:bound_max}) with respect to $\delta$ is equivalent to finding: 
\begin{equation}
\label{eq:boundop}
\begin{aligned}
\max_{x} \quad & x^2\left(1-x\right)
\end{aligned}
\end{equation}
(\ref{eq:boundop}) attains it's maximum value at $x=2/3$. This gives us that: 
\begin{equation}
    \mathcal{M}(C) \geq  \frac{32\pi^2\lambda^2\sigma^2D^4}{27K^2 \left\lVert X^{\mathrm{tag}}\right\rVert_F^2 \sum_{t=1}^T \lVert \widetilde{B}_t\rVert_F^2\left\lVert r(c_t)\right\rVert_2^2}
\end{equation}
\subsection{Proof of Lemma \ref{lem:ugly}}
For ease of computation, we drop the $t$ subscript for now, and define $\widetilde{B}= (I-BR_t)^{-1}$ with $\tilde{b}_{i,j} = \tilde{B}_{i,j}$, and $\tilde{H}_Q = H_Q\tilde{B}$ with $\tilde{h}_{Q,i,j} = (\tilde{H}_Q)_{i,j}$. Taking $\mathbf{s}_t = \mathbf{1}$, we have that:
\begin{align}
\left \lVert A_t \right \rVert_{F}^{2} &= \sum_{k = 1}^{K}\sum_{n = 1}^{N} \left|\sum_{i = 1}^{K} \tilde{h}_{k,n}^{Q} h_{i,n}^{Q} - \tilde{h}_{k,n}^{Q'} h_{i,n}^{Q'} \right|^{2}, \\
\label{eq: A_bound}
&\leq \sum_{k = 1}^{K}\sum_{n = 1}^{N}\sum_{i = 1}^{K} \left|\tilde{h}_{k,n}^{Q} h_{i,n}^{Q} - \tilde{h}_{k,n}^{Q'} h_{i,n}^{Q'} \right|^{2}.
\end{align}
Using Cauchy-Schwartz inequality, we arrive at:  
\begin{align}
    \left|\tilde{h}_{k,n}^{Q_1} h_{i,n}^{Q_1} - \tilde{h}_{k,n}^{Q_2} h_{i,n}^{Q_2} \right|^{2} &\leq \left \lVert {\tilde{\mathbf{b}}}_{n} \right \rVert_{2}^{2}\sum_{l = 1}^{N} \left|h_{k,l}^{Q_1} h_{i,n}^{Q_1} - h_{k,l}^{Q_2} h_{i,n}^{Q_2} \right|^{2}
\end{align}
Assuming we operate in the far-field region, we can approximate $h$ with 
\begin{equation}
    h^{Q}_{i,j} = \frac{e^{-j \left(\frac{2 \pi}{\lambda} \right) \left \Vert \mathbf{x}_{i}^{ant} - Q \mathbf{x}_{j}^{tag} \right \Vert_{2}}}{4 \pi \left \Vert \mathbf{x}_{i}^{ant} - Q\mathbf{x}_{j}^{tag} \right \Vert_{2}} \approx \frac{e^{-j \left(\frac{2 \pi}{\lambda} \right) \left \Vert \mathbf{x}_{i}^{ant} - Q\mathbf{x}_{j}^{tag} \right \Vert_{2}}}{4 \pi D},
\end{equation}
where $D$ is an approximate range between the tagged object's hinge point and the aperture. Hence, we can make the following approximation:
\begin{multline}
 \label{eq:lower13}
 \left|h_{k,l}^Q h_{i,n}^Q - h_{k,l}^{Q'} h_{i,n}^{Q'}  \right|^{2} \approx \\
 \left(\frac{1}{4 \pi D} \right)^{4} \left\lvert e^{-j \left(\frac{2 \pi}{\lambda} \right) \big(\left \Vert \mathbf{x}_{k}^{ant} - Q\mathbf{x}_{l}^{tag} \right \Vert_{2} + \left \Vert \mathbf{x}_{i}^{ant} - Q\mathbf{x}_{n}^{tag} \right \Vert_{2} \big)} \right. \\
    \left. -e^{-j \left(\frac{2 \pi}{\lambda} \right) \big(\left \Vert \mathbf{x}_{k}^{ant} - Q'\mathbf{x}_{l}^{tag} \right \Vert_{2} + \left \Vert \mathbf{x}_{i}^{ant} - Q'\mathbf{x}_{n}^{tag} \right \Vert_{2} \big)}  \right\rvert. 
\end{multline}
Using $1 - e^{-x} \leq x$, we can further upper bound (\ref{eq:lower13}) by: 
\begin{multline}
\label{eq:lower15}
    \left(\frac{\sqrt{\frac{2 \pi}{\lambda}}}{4 \pi D} \right)^{4} \left( \left\lVert \mathbf{x}_{k}^{ant} - Q\mathbf{x}_{l}^{tag} \right\rVert_{2} + \left \lVert \mathbf{x}_{i}^{ant} - Q \mathbf{x}_{n}^{tag} \right\rVert_{2}\right. \\ 
    - \left. \left \lVert \mathbf{x}_{k}^{ant} - Q'\mathbf{x}_{l}^{tag} \right \rVert_{2} - \left \lVert \mathbf{x}_{i}^{ant} - Q'\mathbf{x}_{n}^{tag} \right \rVert_{2} \right). 
\end{multline}
Using the triangle and matrix norm inequalities in (\ref{eq:lower15}), we obtain
\begin{multline}
    \label{eq:upper16}
    \left\lvert h_{k,l}^{Q} h_{i,n}^{Q} - h_{k,l}^{Q'} h_{i,n}^{Q'} \right\rvert^{2} \\
    \leq 2\left(\frac{\sqrt{\frac{2 \pi}{\lambda}}}{4 \pi D} \right)^{4} \left(\left \Vert x_{l}^{\textrm{tag}}   \right \Vert_{2}^{2} + \left \Vert x_{n}^{\textrm{tag}}   \right \Vert_{2}^2 \right)\theta(Q,Q').
\end{multline}
Combining (\ref{eq:upper16}) and (\ref{eq: A_bound}), and summing over the indices of the tags and antennas, we obtain the desired result.
\subsection{Proof of Theorem \ref{thm:dependence}}
\label{sec:proof_dependence}
\noindent Suppose $C = [\mathbf{c}_1 \cdots \mathbf{c}_T]$ and $C'=[\mathbf{c}'_1 \cdots \mathbf{c}'_T]$ are two coding matrices with the same relative weight vectors. In other words, $\mathbf{\pi}_{\mathbf{c}} = \mathbf{\pi}_{\mathbf{c}}'$ for every code configuration $\mathbf{c}$. If the counts of the different configurations are the same, then there exists some permutation $\tau$ on $\{1,\ldots, T\}$ such that $\mathbf{c}_{\tau(t)} = \mathbf{c}_t'$ for all $t$. We show the errors do not depend on the ordering of the code configurations due to the iid nature of the noise. To see this, let $Y = F(Q;C) + W$, then 
\begin{align}
\label{eq:c_hat}
    &\widehat{Q}(Y) \\
    &= \widehat{Q}\left(F(Q;C) + W\right),\\
    &= \arg\min_{Q'} \left\lVert F(Q;C) - F(Q';C) + W\right\rVert_F^2, \\
    &= \arg\min_{Q'} \sum_{t=1}^{T} \left\lVert f(Q;\mathbf{c}_t) - f(Q';\mathbf{c}_t) + \mathbf{w}_t \right\rVert^2, \\
    &= \arg\min_{Q'} \sum_{t=1}^{T} \left\lVert f(Q;\mathbf{c}_\tau(t)) - f(Q';\mathbf{c}_\tau(t)) + \mathbf{w}_{\tau(t)} \right\rVert^2, \\
    &= \arg\min_{Q'} \sum_{t=1}^{T} \left\lVert f(Q;\mathbf{c}_t') - f(Q';\mathbf{c}_t') + \mathbf{w}_{\tau(t)} \right\rVert^2, \\
     &=  \widehat{Q}\left(F(Q;C') + W'\right), \\
    \label{eq:c_hatt}
     &= \widehat{Q}(Y'),
\end{align}
where $\mathbf{w}_t$ is the $t^{\mathrm{th}}$ $K$-sized block of $W$, $W'$ is the vector with $\mathbf{w}'_t = \mathbf{w}_{\tau(t)}$, and $Y' = F(Q;C') + W'$. Since the elements of $W$ independent and identically distributed, we get that $W$ and $W'$ are also identically distributed. Combining the previous statement with the fact that $W$ and $W'$ are permutations of one another, we can replace any expectation with respect to $W$ with an expectation with respect to $W'$. Hence, 
\begin{align}
    \mathcal{L}(C) &= \sum_{Q\in \mathcal{Q}} \mathbb{E}_{W} \left[\theta(\widehat{Q}(Y), Q)\right], \\
    \label{eq:step_1}
    &= \sum_{Q\in \mathcal{Q}} \mathbb{E}_{W} \left[\theta(\widehat{Q}(Y'), Q)\right], \\
    \label{eq:step_2}
    &= \sum_{Q\in \mathcal{Q}} \mathbb{E}_{W'} \left[\theta(\widehat{Q}(Y'), Q)\right], \\
    &=\mathcal{L}(C').
\end{align}
(\ref{eq:step_1}) follows from the earlier computation, and (\ref{eq:step_2}) follows from the our earlier discussion. Repeating the exact same computation with $\sum_{Q\in\mathcal{Q}}$ replaced with $\max_{Q\in\mathcal{Q}}$ shows that $\mathcal{M}(C) = \mathcal{M}(C')$ i.e. $C$ and $C'$ share the same worst-case performance. Hence, we get that $C$ and $C'$ share the same errors, and that our performance measures depend only on the coding matrix through the relative weight vector. It is worthwhile to note that this property of the errors is not unique to Gaussian noise, but is valid for any noise vector whose components are independent and identically distributed. 
\section{Conclusion}
Active backscatter tags can be very useful when estimating the 3D orientation of objects if tag responses are carefully designed. In this work, we suggested two design criteria, and developed a bound showing how different system parameters affect performance. Through numerical simulations, we exhibited the effectiveness of the suggested designs, and their robustness in the face of imperfect channel knowledge. There are several open questions remaining including designs for more complicated or time-varying environments. Additionally, an exploration of the effectiveness of non-coherent coding in such scenarios is a potentially interesting direction. 
\bibliographystyle{IEEEtran}
\bibliography{bibtex}
\end{document}

%% file: table.tex
  \resizebox{5.15cm}{!}{
  \begin{tabular}{|c|cccc|}
  \hline
  \diagbox[width=6.5em, height=2.5em]{Tag}{Time} & 1 & 2 & $\cdots$ & T \\ \hline
  1 & \begin{tikzpicture}[baseline=0]
    \node[draw, fill=blue(munsell), star,star points=7,star point ratio=0.8, inner sep=2.5] {};    \addvmargin{1mm}
  \end{tikzpicture}   &    \begin{tikzpicture}[baseline=0]
    \node[draw, fill=blue(munsell), star,star points=7,star point ratio=0.8, inner sep=2.5] {};    \addvmargin{1mm}
  \end{tikzpicture}   & $\cdots$ & \begin{tikzpicture}[baseline=0]
    \node[draw, fill=amber, star,star points=7,star point ratio=0.8, inner sep=2.5] {};    \addvmargin{1mm} 
  \end{tikzpicture} \\ 
  2 & \begin{tikzpicture}[baseline=0]
    \node[draw, fill=amber, star,star points=7,star point ratio=0.8, inner sep=2.5] {};    \addvmargin{1mm}
  \end{tikzpicture}   &    \begin{tikzpicture}[baseline=0]
    \node[draw, fill=amber, star,star points=7,star point ratio=0.8, inner sep=2.5] {};    \addvmargin{1mm}
  \end{tikzpicture}   & $\cdots$ & \begin{tikzpicture}[baseline=0]
    \node[draw, fill=blue(munsell), star,star points=7,star point ratio=0.8, inner sep=2.5] {};    \addvmargin{1mm} 
  \end{tikzpicture} \\ 
  \vdots & \vdots  &    \vdots   & $\ddots$ & \vdots \\
  N & \begin{tikzpicture}[baseline=0]
    \node[draw, fill=blue(munsell), star,star points=7,star point ratio=0.8, inner sep=2.5] {};    \addvmargin{1mm}
  \end{tikzpicture}   &    \begin{tikzpicture}[baseline=0]
    \node[draw, fill=amber, star,star points=7,star point ratio=0.8, inner sep=2.5] {};    \addvmargin{1mm}
  \end{tikzpicture}   & $\cdots$ & \begin{tikzpicture}[baseline=0]
    \node[draw, fill=amber, star,star points=7,star point ratio=0.8, inner sep=2.5] {};    \addvmargin{1mm} 
  \end{tikzpicture} \\ \hline
  \end{tabular}
  }

%% file: matrix.tikz
\begin{tikzpicture}[baseline={-0.5ex},mymatrixenv, scale=1.5]
\matrix [mymatrix,inner sep=5.5pt] (m) at (7,-1.5){ 
     1&1&\cdots&0\\
     0&0&\cdots&1\\
     \vdots&\vdots&\ddots&\vdots\\
     1&0&\cdots&0\\
  };

 \begin{scope}[on background layer,rounded corners]
     \node [fit=(m-1-1) (m-1-4),greenish,inner xsep=0pt,inner ysep=2.5pt]{};
     \node [fit=(m-4-1) (m-4-4),redish,inner xsep=0pt,inner ysep=2.5pt]{};
     \node [fit=(m-1-1) (m-4-1),purpleish,inner xsep=0.5pt,inner ysep=3.5pt]{};
     \node [fit=(m-1-2) (m-4-2),orangeish,inner xsep=0.5pt,inner ysep=3.5pt]{};
     \node [fit=(m-1-4) (m-4-4),cyanish,inner xsep=0.5pt,inner ysep=3.5pt]{};
    \end{scope}
\begin{scope}[every node/.append style={scale=\myscale,transform
    shape},very thick]
        \mymatrixbracebottom{1}{1}{ \footnotesize$\mathbf{c}_1$}
        \mymatrixbracebottom{2}{2}{ \footnotesize$\mathbf{c}_2$}
        \mymatrixbracebottom{4}{4}{ \footnotesize$\mathbf{c}_T$}
        \mymatrixbraceleft{1}{4}{\begin{tabular}{l}
    \footnotesize Tag $1$\\
    \footnotesize States 
\end{tabular}}
\mymatrixbraceleft{4}{4}{\begin{tabular}{l}
    \footnotesize Tag $N$ \\
    \footnotesize States 
\end{tabular}}
    \end{scope} 
\end{tikzpicture}